\documentclass[%
reprint,
superscriptaddress,
altaffilletter,
showpacs,
showkeys,
amsmath, amssymb, aps,prl, floatfix,
]{revtex4-1}

\usepackage{graphicx}% Include figure files
\usepackage{dcolumn}% Align table columns on decimal point
\usepackage{bm}% bold math
\usepackage{comment}% commenting blocks
\usepackage[usenames,dvipsnames]{color}
\usepackage[dvipsnames]{xcolor}
\usepackage[normalem]{ulem}
\usepackage[symbol]{footmisc}
\usepackage{xspace}
\usepackage{ulem}
\usepackage{hyperref}% add hypertext capabilities
\hypersetup{colorlinks = true, allcolors = blue}
\usepackage{multirow}
\DeclareGraphicsExtensions{.pdf,.png,.jpg}
\newcommand{\SeMass}{4.65~kg\xspace}

\newcommand{\startDAQ}{June 2017}

\newcommand{\mvveffred}{0.0760~^{+~0.0006}_{-~0.0007}}

\newcommand{\HalfLifeold}{T_{1/2}^{2\nu} = [8.60 \pm 0.03 \textrm{(stat.)}~^{+0.19}_{-0.13} \textrm{(syst.)}] \times 10^{19}~\textrm{yr}}

\newcommand{\HalfLifenewred}{T_{1/2}^{2\nu} = [8.69 \pm 0.05 \textrm{(stat.)}~^{+0.09}_{-0.06} \textrm{(syst.)}] \times 10^{19}~\textrm{yr}}

\newcommand{\BIoneexp}{$(3.5 \pm 1.0)\times 10^{-3}$}
\newcommand{\BItwoexp}{$(5.5 \pm 1.5)\times 10^{-3}$}
\newcommand{\BIonemodel}{$[4.2 \pm 0.2(stat.) \pm 0.4(syst.)]\times 10^{-3}$}
\newcommand{\BItwomodel}{$[4.0 \pm 0.2(stat.) _{-0.2}^{+0.4}(syst.)]\times 10^{-3}$}
\newcommand{\vvactivitySSD}{[$8.63 \pm 0.04$ (stat.)]~mBq}
\newcommand{\ZS}{Zn$^{82}$Se\xspace}
\newcommand{\se}{$^{82}$Se\xspace}
\newcommand{\mo}{$^{100}$Mo\xspace}

\newcommand{\onu}{$0\nu\beta \beta$\xspace}

\newcommand{\TL}{$^{208}\mathrm{Tl}$\xspace}

\newcommand{\KF}{$^{40}\mathrm{K}$\xspace}
\newcommand{\THO}{$^{232}\mathrm{Th}$\xspace}
\newcommand{\UR}{$^{238}\mathrm{U}$\xspace}
\newcommand{\URA}{$^{235}\mathrm{U}$\xspace}

\newcommand{\PO}{$^{210}\mathrm{Po}$\xspace}

\newcommand{\SR}{$^{90}$Sr/$^{90}$Y\xspace}

\newcommand{\K}{$^{40}\mathrm{K}$\xspace}

\newcommand{\QBB}{$\mathrm{Q}_{\beta\beta}$\xspace}

\newcommand{\vv}{2$\nu\beta\beta$\xspace}
\newcommand{\ckky}{counts/keV/kg/yr\xspace}

\newcommand{\mspeconea}{$\mathcal{M}^{\alpha}_{1}$\xspace}
\newcommand{\mspeconeb}{$\mathcal{M}^{\beta/\gamma}_{1}$\xspace}
\newcommand{\mspectwo}{$\mathcal{M}_2$\xspace}

\newcommand{\summspectwo}{$\Sigma_2$\xspace}
\newcommand{\phone}{phase-I\xspace}
\newcommand{\phtwo}{phase-II\xspace}

\begin{document}

\title{Measurement of the \vv Decay Half-Life of Se-82 with the Global CUPID-0 Background Model}

% All university affiliations addresses go here:
\newcommand{\sapienza}{\affiliation{Dipartimento di Fisica, Sapienza Universit\`a di Roma, P.le Aldo Moro 2, 00185, Roma, Italy}}
\newcommand{\infnroma}{\affiliation{INFN, Sezione di Roma, P.le Aldo Moro 2, 00185, Roma, Italy}}
\newcommand{\cnr}{\affiliation{Consiglio Nazionale delle Ricerche, Istituto di Nanotecnologia, c/o Dip. Fisica, Sapienza Università di Roma, 00185, Rome, Italy}}
\newcommand{\lnl}{\affiliation{INFN  Laboratori Nazionali di Legnaro, I-35020 Legnaro (Pd) - Italy}}
\newcommand{\lngs}{\affiliation{INFN  Laboratori Nazionali del Gran Sasso, I-67100 Assergi (AQ) - Italy}}
\newcommand{\lbl}{\affiliation{Lawrence Berkeley National Laboratory , Berkeley, California 94720, USA}}
\newcommand{\infnge}{\affiliation{INFN  Sezione di Genova, I-16146 Genova - Italy}}
\newcommand{\unige}{\affiliation{Dipartimento di Fisica, Universit\`{a} di Genova, I-16146 Genova - Italy}}
\newcommand{\infnmib}{\affiliation{INFN  Sezione di Milano - Bicocca, I-20126 Milano - Italy}}
\newcommand{\unimib}{\affiliation{Dipartimento di Fisica, Universit\`{a} di Milano - Bicocca, I-20126 Milano - Italy}}
\newcommand{\csnsm}{\affiliation{CNRS/CSNSM, Centre de Sciences Nucl$\acute{e}$aires et de Sciences de la Mati$\grave{e}$re, 91405 Orsay, France}}
\newcommand{\cea}{\affiliation{IRFU, CEA, Universit$\acute{e}$ Paris-Saclay, F-91191 Gif-sur-Yvette, France}}
\newcommand{\gssi}{\affiliation{Gran Sasso Science Institute, 67100, L'Aquila - Italy}}
\newcommand{\usc}{\affiliation{Department of Physics  and Astronomy, University of South Carolina, Columbia, SC 29208 - USA}}
\newcommand{\mpi}{\affiliation{Max-Planck-Institut fÃŒr Physik, D-80805 MÃŒnchen, Germany}}
\newcommand{\dis}{\affiliation{DISAT, Universit\`a dell'Insubria, 22100 Como, Italy}}
\newcommand{\JYU}{\affiliation{University of Jyv\"askyl\"a, Department of Physics, P. O. Box 35 (YFL), FI-40014, Finland}}
\newcommand{\FIER}{\affiliation{Finnish Institute for Educational Research, P.O.Box 35 FI-40014 University of  Jyv\"askyl\"a - Finland}}
\newcommand{\CTP}{\affiliation{Center for Theoretical Physics, Sloane Physics Laboratory, Yale University, New Haven, Connecticut 06520-8120 - USA}}
\newcommand{\QUEEN}{\affiliation{Department of Physics, Engineering Physics and Astronomy, Queen's University, K7L 3N6 Kingston, ON, Canada}}
\newcommand{\mcdonald}{\affiliation{Arthur B. McDonald Canadian Astroparticle Physics Research Institute, K7L 3N6 Kingston, ON, Canada}}

\author{O.~Azzolini}\lnl
\author{J.~W.~Beeman}\lbl
\author{F.~Bellini}\sapienza\infnroma
\author{M.~Beretta}\altaffiliation{Present address: Department of Physics, University of California, Berkeley, CA 94720, USA}\unimib\infnmib
\author{M.~Biassoni}\infnmib
\author{C.~Brofferio}\unimib\infnmib
\author{C.~Bucci} \lngs
\author{S.~Capelli}\unimib\infnmib
\author{V.~Caracciolo}\altaffiliation{Present address: Dipartimento di Fisica, Universit\`{a} di Roma Tor Vergata, I-00133, Rome, Italy }\lngs
\author{L.~Cardani}\infnroma
\author{P.~Carniti}\unimib\infnmib
\author{N.~Casali}\infnroma
\author{E.~Celi}\email[Corresponding author: ]{emanuela.celi@gssi.it}\gssi\lngs
\author{D.~Chiesa}\unimib\infnmib
\author{M.~Clemenza}\unimib\infnmib
\author{I.~Colantoni}\infnroma\cnr
\author{O.~Cremonesi}\infnmib
\author{A.~Cruciani}\infnroma
\author{A.~D'Addabbo}\lngs
\author{I.~Dafinei}\gssi\infnroma
\author{S.~Di~Domizio}\unige\infnge
\author{V.~Dompè}\sapienza\infnroma
\author{G.~Fantini}\sapienza\infnroma
\author{F.~Ferroni}\infnroma\gssi
\author{L.~Gironi}\unimib\infnmib
\author{A.~Giuliani}\csnsm
\author{P.~Gorla}\lngs
\author{C.~Gotti}\infnmib
\author{G.~Keppel}\lnl
\author{J.~Kotila}\JYU\FIER\CTP
\author{M.~Martinez}\altaffiliation{Present address: Centro de Astropart\'iculas y Física de Altas Energ\'ias, Universidad de Zaragoza, and ARAID, Fundaci\'on Agencia Aragonesa para la Investigaci\'on y el Desarrollo, Gobierno de Arag\'on, Zaragoza 50018, Spain}\sapienza\infnroma
\author{S.~S.~Nagorny}\altaffiliation{Present address: Department of Physics $\&$ Engineering Physics Astronomy, Queen's University Kingston, Ontario, K7L 3N6 Kingston, Canada}\lngs
\author{M.~Nastasi}\unimib\infnmib
\author{S.~Nisi}\lngs
\author{C.~Nones}\cea
\author{D.~Orlandi}\lngs
\author{L.~Pagnanini}\email[Corresponding author: ]{lorenzo.pagnanini@gssi.it}\gssi\lngs
\author{M.~Pallavicini}\unige\infnge
\author{L.~Pattavina}\altaffiliation{Present address: Physik-Department and Excellence Cluster Origins, Technische Universit{\"a}t M{\"u}nchen, 85747 Garching, Germany}\lngs
\author{M.~Pavan}\unimib\infnmib
\author{G.~Pessina}\infnmib
\author{V.~Pettinacci}\infnroma
\author{S.~Pirro}\lngs
\author{S.~Pozzi}\unimib\infnmib
\author{E.~Previtali}\unimib\lngs
\author{A.~Puiu}\lngs
\author{A.~Ressa}\sapienza\infnroma
\author{C.~Rusconi}\lngs\usc
\author{K.~Sch\"affner}\altaffiliation{Present address: Max-Planck-Institut f{\"u}r Physik, 80805 M{\"u}nchen - Germany}\lngs
\author{C.~Tomei}\infnroma
\author{M.~Vignati}\sapienza\infnroma
\author{A.~S.~Zolotarova}\cea

\date{\today}

\begin{abstract}
We report on the results obtained with the global CUPID-0 background model, which combines the data collected in the two measurement campaigns for a total exposure of 8.82~kg$\times$yr of $^{82}$Se. 
We identify with improved precision the background sources within the 3 MeV energy region, where neutrinoless double $\beta$-decay of \se and \mo is expected, making more solid the foundations for the background budget of the next-generation CUPID experiment. Relying on the excellent data reconstruction, we measure the two-neutrino double $\beta$-decay half-life of $^{82}$Se with unprecedented accuracy:
$\HalfLifenewred$.
\end{abstract}

\pacs{07.20.Mc, 23.40.-s, 21.10.Tg, 27.50.+e}
\keywords{scintillating bolometers, background model, two-neutrino double $\beta$-decay, spectral shape}
\maketitle
%***********************Introduction
{\it Introduction.} The comprehension of the neutrino nature is of fundamental importance to understand the origin of neutrino masses and to have hints of new physics in lepton sector. \textcolor{black}{The Standard Model~(SM) predicts neutrino to be Dirac particle ($\nu \neq\bar{\nu}$), and the neutrinoless double $\beta$-decay (\onu)~\cite{Furry:1939qr} is the only practical way to probe if neutrinos are Majorana particles ($\nu=\bar{\nu}$)~\cite{DellOro:2016tmg}.} The double $\beta$-decay is a weak process involving the decay of two neutrons into protons with the emission of two electrons. According to the SM, such a process is allowed only with the emission of two anti-neutrinos in the final state~\cite{GoeppertMayer}, the so-called two-neutrino double $\beta$-decay (\vv). In the past decades, big experimental efforts were committed to the search for \onu allowing the existing detectors to reach sensitivities of the order of $10^{25}-10^{26}$~yr on its half-life~\cite{GERDA:2020xhi,KamLAND-Zen:2022tow,CUORE:2021mvw}. Nowadays, none of them has ever observed such decay, however, the development of technologies able to reach excellent sensitivities led to outstanding precision measurements of \vv~\cite{Saakyan:2013yna,Barabash:2015eza,NEXT:2021dqj,GERDA:2023wbr,KamLAND-Zen:2019imh,CUORE:2020bok,CUPID-Mo:2023lru,NEMO-3:2019gwo}. 
The properties of cryogenic calorimeters make them competitive detectors both in the search for \onu and in the measurement of \vv. Being the signal source embedded into the detector, this technology offers high detection efficiency and excellent energy resolution. Furthermore, using different crystals opens the opportunity to study several double-$\beta$ decay emitters. 
In the past years, the CUORE (Cryogenic Underground Observatory for Rare Events) experiment demonstrated the possibility to reach ton-scale exposures~\cite{CUORE:2021mvw}, while several R\&D projects~\cite{Pirro:2005ar,Arnaboldi:2010jx,Beeman:2013vda,Artusa:2016maw,Armengaud:2017hit} developed a powerful technology based on scintillating bolometers to enable particle discrimination, thus reducing the radioactive background. 
These lay the foundations for CUPID (CUORE Upgrade with Particle IDentification), a next-generation experiment aiming to reach unprecedented sensitivities using a large-size array of scintillating cryogenic calorimeters~\cite{CUPIDInterestGroup:2019inu}.
\textcolor{black}{CUPID-0 is the first demonstrator of the scintillating bolometer technique in the \onu field and it has allowed to optimize important aspects of the CUPID design, such as light detector, reflective foil, and holder structure. CUPID-0 is also able to demonstrate the origin of the background in the Region of the Interest (ROI) of \se \onu, where also the $^{100}$Mo \onu is expected.} In this letter, we present the results of the global background model of CUPID-0, and a novel measurement of the $^{82}$Se-\vv half-life, which results to be the most precise so far.

%***********************Detector 
\begin{figure*}
\includegraphics[width=0.9\textwidth]{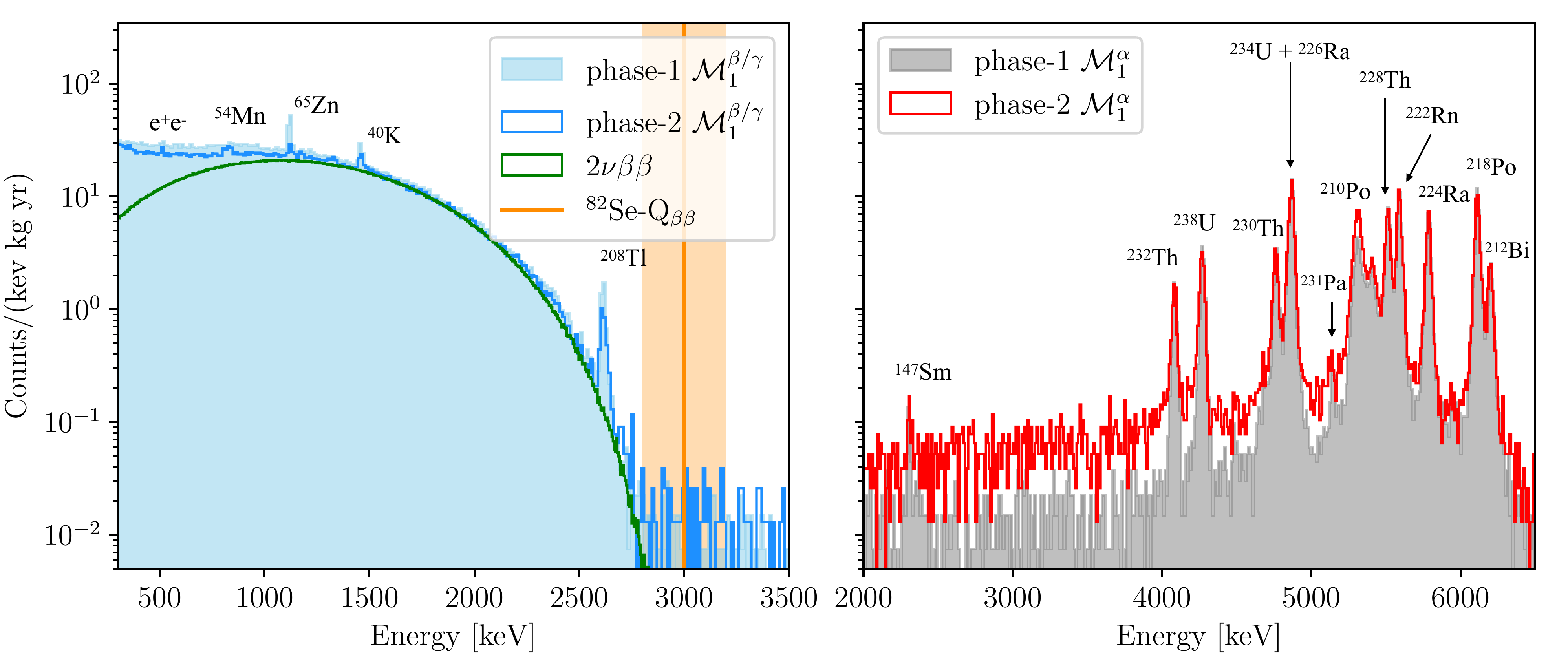}
\caption{\textcolor{black}{\mspeconeb (left) and \mspeconea (right) spectra registered by CUPID-0 in \phone (full histograms) and \phtwo (line histograms). In \mspeconeb, we identify the dominant \vv contribution (green), and the $\gamma$-rays peaks due to cosmogenic activation of \ZS ($^{65}$Zn), and setup contaminants (\KF,\TL). In \mspeconea, degraded $\alpha$-particles produce a flat background at 2-4~MeV, while the $\alpha$-peaks above are the distinctive signatures of the different detector contaminants. The main differences between \phone and \phtwo traceable in the data are the $^{65}$Zn reduction due to its short half-life, the \KF and \TL reduction due to the 10~mK (1 cm of copper) shield, which also introduces an additional \PO component in the 2-4~MeV region of \mspeconea. The yellow band in \mspeconeb marks the Region of Interest of \se \onu.}}
\label{fig:spectra}
\end{figure*}

{\it CUPID-0 detector.} CUPID-0 is composed of 24~ZnSe scintillating crystals~\cite{Dafinei:2017xpc} enriched at (95.3~$\pm$~0.3)\% in $^{82}$Se~\cite{Balabanov:2023tnj}.
The array is completed by two natural ZnSe crystals, for a total ZnSe mass of 10.5~kg. 
The scintillation light is detected by thin Germanium wafers operating as cryogenic calorimeters~\cite{Beeman:2013zva}.
A Vikuiti$^{\text{TM}}$ plastic reflective foil surrounds each ZnSe crystal to enhance the light collection. 
ZnSe crystals and light detectors are alternatively stacked in 5 towers, and supported by a copper structure to which they are connected by means of PTFE clamps. 
To convert the temperature variation produced by interacting particles into a voltage signal, a germanium thermistor is glued to each device (both light detectors and ZnSe) and bonded to the cryogenic readout system through thin gold wires. 
The detector is installed in an Oxford~1000~$^3$He/$^4$He dilution refrigerator working at a base temperature of 12~mK. The whole experimental setup is located underground in the Laboratori Nazionali del Gran Sasso (LNGS), in Italy. 
More details on the CUPID-0 detector setup can be found in Refs.~\cite{Azzolini:2018tum,Carniti:2017zkr,Arnaboldi:2015wvc,Arnaboldi:2017aek}. The entire data-taking amounts to nearly three years, from \startDAQ~to February~2020, with a few months off from January to June~2019 to upgrade the detector. We refer to the two physics runs of the experiment as \phone (9.95~kg$\times$yr of \ZS) and \phtwo (5.74~kg$\times$yr of \ZS), respectively.
In the \phtwo, the reflective foils are removed and we add a further copper shield at the 10~mK stage directly facing the detector. In \phone, the reflectors between the ZnSe crystals affect the correct identification of surface/bulk $\alpha$-decays due to detector contaminants~\cite{Azzolini:2021yft}. Moreover, from the \phone background model, we also noticed a possible background affecting the \onu region of interest from the elements near the detector. 
Excluding the reflectors from the possible background sources is fundamental to assess if they can be used in next-generation experiments~\cite{CUPID:2020itw}.
Without the reflecting foils, the $\alpha$ discrimination power is still excellent~\cite{CUPID:2022puj}, and we can disentangle contaminants coming from crystal surfaces and components close to the detector. 
Moreover, the presence of the shield at 10~mK allows us to better study the cryostat contaminants, exploiting the different counting rates of the $\gamma$-ray peaks in \phone and \phtwo (see Fig.\ref{fig:spectra}).

%*********************** Data taking and analysis.
{\it Data taking and analysis.} Given an energy release inside the ZnSe, the relative voltage signal is triggered and saved in a 5~s time window. The output of the corresponding light detectors is recorded by analyzing the coincidence in time with the ZnSe signal. Light pulses are acquired in a 500~ms time window~\cite{DiDomizio:2018ldc}.
We apply a matched-filter algorithm~\cite{Gatti:1986cw} on the raw data to estimate the pulse height and the shape parameters. Then, we calibrate each detector by analyzing the data acquired with an external $^{232}$Th source. The $^{232}$Th decay chain provides several $\gamma$-lines in the range (511-2615)~keV that can be used as reference peaks by fitting with a zero intercept parabolic function the non-calibrated amplitudes. A dedicated $^{56}$Co calibration with $\gamma$-lines above 2615~keV is also performed to investigate the uncertainty on the energy scale with the distribution of residuals~\cite{Azzolini:2019tta}. Finally, we tag the $\alpha$ events by looking at the pulse shape parameters of the light signal pulses~\cite{Azzolini:2018yye}.

Different sources can be disentangled thanks to the following features: (i) the identification of $\alpha$ particles from $\beta$/$\gamma$ down to 2 MeV; (ii) the excellent resolution, enabling the study of the distortion in the shape of $\alpha$ peaks, essential to tag the surface contaminants; (iii) the granularity of the detector, exploited to distinguish interactions in which one or multiple crystals are involved. For this purpose, we tag the events triggering multiple crystals within a $\pm20$~ms time window.

In this analysis, we refer to the \mspeconeb and \mspeconea spectra as the single-hits events of $\beta$/$\gamma$ and $\alpha$ respectively (see (see Fig.\ref{fig:spectra}), while to \mspectwo and \summspectwo spectra as double-hit events (\mspectwo comprises the energies detected by each crystal, \summspectwo the total energy released in two crystals). Data selection is based on several quality cuts to reject non-physical events and multiple pulses in the same acquisition window. 
The relative efficiency of these cuts is evaluated for \phone and \phtwo separately and combined with the trigger and energy reconstruction efficiencies~\cite{Azzolini:2019nmi}. 
The result is $\varepsilon_{1}=(95.7 \pm 0.5)$ and $\varepsilon_{2}=(94.8 \pm 0.7)$ for \phone and \phtwo, respectively.

%***********************Modeling (MC)
{\it Data Modeling.} Initially, we reproduce the CUPID-0 geometry with the \textsc{Geant4} toolkit~\cite{Geant4}. 
Following the same grouping criteria used in the \phone background model, we split the detector components in \textit{Crystals}, \textit{Holder}, \textit{Reflectors} (\phone), and \textit{10mK} (\phtwo).
The light detectors are included in the geometry but not considered as a background source, since their radioactive contaminants are negligible~\cite{Azzolini:2019nmi}.
The groups named \textit{Holder}, \textit{Reflectors}, and \textit{10mK} include also the elements close to the detector (e.g. wiring, glue), whose contaminants will not be generated.
The different components of shielding and dilution refrigerator are also reproduced in detail, and grouped under the label \textit{Cryostat} in the following.
We separately generate for \phone and \phtwo the radioactive decays from various background sources in each group of volumes.
In particular, we simulate the initial kinematics of the \vv assuming the Single State Dominance~(SSD) mechanism, calculating the electron energy using exact Dirac electron wave functions that take into account finite nuclear size and electron screening~\cite{Kotila:2012zza}, while the generation of external muons at LNGS described in detail in Ref.~\cite{CUORICINO:2009zeh}.
Each simulation is then processed with a custom software to reproduce both detector and data processing features, like time and energy resolution, energy threshold, $\alpha$-particle identification, and time coincidence window~\cite{Azzolini:2018yye}.
The identification of the radioactive sources and their location is driven by material screening results and CUPID-0 data analysis. An $\alpha$-decay occurring in crystal bulk produces a peak at the Q-value of the decay in \mspeconea, while contaminants on the crystal surface can be distinguished in \mspectwo and \summspectwo. In \phone, the presence of reflecting foils 
suppress the signature of surface contaminations due to the absorption of $\alpha$-particles or recoil escapes. However, in this analysis, we can recover that information by combining it with the \phtwo data. \THO and \UR decay chains in the close components can produce sharp peaks at the nominal $\alpha$-particle energies if the contaminations are very shallow, or a continuum spectrum if the decays occur deeply from the surface.
To model the surface contaminants both for crystals and close components (\textit{Reflectors}, \textit{Holder}, \textit{10mK}), we parameterize its density profile as $e^{-x/\lambda}$, where $\lambda$ is the depth parameter. Based on previous experiences~\cite{Azzolini:2019nmi}, we used $\lambda = 10$~nm to model shallow contamination and $\lambda = 10$~$\mu$m for deeper ones.
Besides this, the $\gamma$-ray lines in \mspeconeb, \mspectwo, and \summspectwo are typically the signature of the external sources (e.g. Cryostat and shields). In this regard, comparing the spectra of \phtwo (additional $1$ cm thick \textit{10mK} shield) with those of \phone, allow us to improve the knowledge of the cryostat contaminants exploiting the relative reduction of the $\gamma$-ray peaks.
For these sources, we also exploited the information from the background model of the CUORE-0 experiment that was hosted in the same facility~\cite{Alduino:2016vtd,Azzolini:2019nmi}.
In summary, we included as bulk contamination the natural decay chains due to \THO, \UR, \URA, and \K for all the simulated elements, the cosmogenic activation products of copper ($^{54}$Mn, $^{58}$Co, $^{60}$Co) and ZnSe ($^{65}$Zn, $^{60}$Co), and the \se-\vv decay. 
Moreover, we included the \THO and \UR contaminations on the surfaces of ZnSe crystals and close detector components. Being the environmental $\gamma$-rays and the neutrons reduced to a negligible level by dedicated shielding~\cite{Alduino:2016vtd}, we consider only muons as an external source. We identify the breakpoints of the secular equilibrium in the radioactive decay chains at $^{228}$Ra for the \THO chain, and at $^{226}$Ra/$^{210}$Pb for the \UR one, as expected from the different chemical behavior. Finally, the muon contribution is normalized on the number of shower events that simultaneously trigger more than three crystals. The result of such normalization is compatible with the one obtainable using the muon flux at LNGS \cite{Ambrosio:1995cx}.

\begin{figure}[!t]
\begin{center}
\includegraphics[width=.48\textwidth]{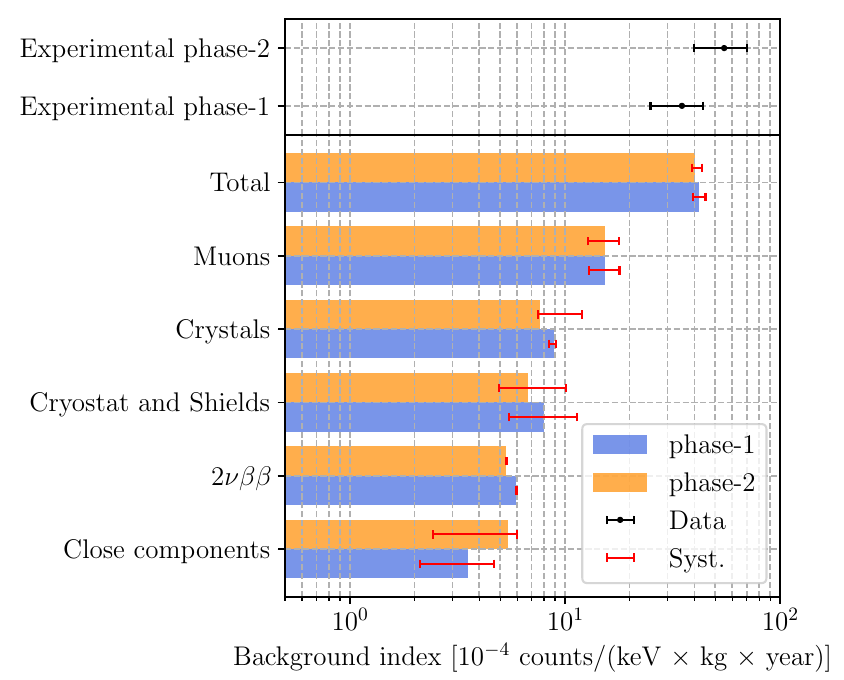}
\caption{\textcolor{black}{Background index in the region of interest for the \onu of \se, i.e. [2.8,3.2]~MeV of the \mspeconeb spectrum.} Experimental background indexes (black, top pad) are compared to the reconstructed ones for \phone~(blue) and \phtwo~(orange). The total background index is the sum of all components with their correlations, and error bars represent the systematic uncertainty of the background reconstruction.}
\label{fig:ROI}
\end{center}
\end{figure}

{\it Spectral Fit.} Exploiting a multivariate Bayesian fit, we reproduce the experimental spectra with a linear combination of the simulated spectra of the different background sources. 
As a result, we obtain a reliable estimation of the activity of each source considered. The fit is performed with the JAGS software~\cite{JAGS}, which exploits the Monte Carlo Markov Chain (MCMC) to sample the joint posterior probability density function of the model unknowns, which linearly depend on the source activities.
We performed a combined fit on 8 experimental spectra, corresponding to \mspeconeb, \mspeconea, \mspectwo, and \summspectwo for both \phone and \phtwo. The fit energy range extends from 300~keV to 5~MeV for \mspeconeb, and from 2 to 7~MeV for \mspeconea. For the \mspectwo and \summspectwo, we require that both events deposit an energy higher than 150~keV, with the exception of events in time coincidence with $\alpha$-particles, for which the threshold is 100~keV to include the nuclear recoils. We use a non-uniform bin size to compact $\alpha$-peaks and $\gamma$-ray lines in a single bin, avoiding the mismatch between simulations and real data induced by the line shape of the peaks. Moreover, we grouped more bins together if the total number of events is lower than 30 counts, while, in any other case, we use a 15~keV fixed step binning. We use non-negative flat priors for all the sources whose activity cannot be constrained \textit{a priori} with independent measurements. In particular, we use the results obtained in the CUORE-0 background model to set a prior on the \THO and \UR in the \textit{Holder} and the $^{60}$Co in the cryostat components. Since the \phone and \phtwo data sets are consecutive in time, we can reasonably assume that the activity of all the contaminants with a half-life longer than the overall CUPID-0 lifetime is constant. For this reason, we constrain a subset of simulations to have the same activity between \phone and \phtwo.
We have a total of 78 simulations, with 17 couples of them constrained~\cite{Supplemental}. We consider as \textit{Reference Fit} the one performed with the simulations, the energy ranges, and the binning described above. The fit details, like the spectral reconstruction over the 8 experimental spectra and the pull distribution, are shown in Ref.~\cite{Supplemental}.
We reconstruct the background index in the ROI by selecting the events in a 400~keV interval centered at the \QBB of \se-\vv \textcolor{black}{, i.e. [2.8,3.2]~MeV}.
We also apply the delayed coincidence cut on the \THO simulations in order to reproduce the selection criteria of the \onu analysis~\cite{CUPID:2022puj,Azzolini:2019nmi}. 
The background index reconstructed with the fit for each group of sources in \phone and \phtwo is reported in Fig.~\ref{fig:ROI}. 

%Systematics
{\it Systematics.} To evaluate the systematic uncertainty affecting the background index, we repeat the fit multiple times in different conditions. We include in this category different energy thresholds (400, 500, 600, 700~keV) and fixed step binnings (30, 50, and 100~keV) in \mspeconeb. Moreover, we run the fit using a reduced list of sources, removing those sources whose contribution results at limit in the \textit{Reference} fit. We evaluate also systematic effects due to energy miscalibration, re-scaling the data for the residuals calculated in Ref.~\cite{Azzolini:2018yye}.
We probe the uncertainty about the contaminant position in cryostat and shields removing alternatively one of the elements from the source list. We investigate an alternative description of surface crystal contaminants and close components. In the former case, we substitute the simulations at 10~nm depth with the 1~nm equivalent ones, while in the latter case, we substitute the \THO and \UR contamination at 10~$\mu$m with bulk ones in the {\it Reflectors} (\phone).
Furthermore, we perform supplementary fits by switching off the $\alpha$-particle identification and using non-informative priors for all the contributions. Besides, we consider also a possible contamination due to \SR in the ZnSe crystals.
\SR is an anthropogenic isotope that decays purely through $\beta$-emission with a lifetime of 28.8~y and produces a featureless spectrum that correlates with the \vv one in the fit. Since the presence of \SR in the ZnSe crystals is unclear, we consider this possibility as systematic. Finally, we account for possible theoretical uncertainties on the \vv decay model generating the initial kinematic of the electrons with an alternative spectrum calculated under the Closure Approximation~\cite{Kotila:2012zza}.

{\it Background Index.} Ultimately, after the investigation of systematic uncertainties, we reproduce a background index \textcolor{black}{in the [2.8,3.2]~MeV energy range} of \BIonemodel~\ckky for \phone and 
\BItwomodel~\ckky for \phtwo, compatible among them and with the observed background of \BIoneexp~\ckky and 
\BItwoexp~\ckky for \phone and \phtwo respectively.
\begin{figure}[!ht]
\begin{center}
\includegraphics[width=.49\textwidth]{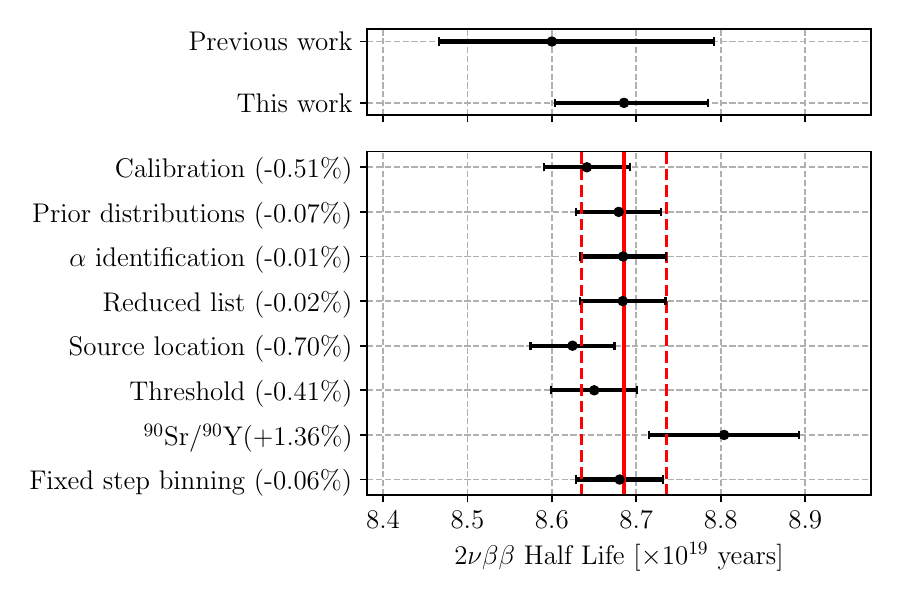}
\caption{Measurement of the \se-\vv half-life obtained in the systematic tests (bottom) and as final result of this analysis, compared to the previous result of CUPID-0 (top). The error bars shown in the bottom plot correspond to the statistical error of each systematic test, while the red vertical lines correspond to the \textit{Reference} mean value (solid line) and its statistical uncertainty (dashed line). The \SR strongly correlates with the \vv, thus broadening the uncertainty. The statistical uncertainty is then combined with the systematic one and compared with the previous result (top).}
\label{fig:2nu}
\end{center}
\end{figure}
According to our model, we can attribute the slightly higher background observed in the ROI of  \se \onu in \phtwo to a statistical fluctuation. \textcolor{black}{Thanks to the combined analysis, we reduce the uncertainty on the \phone background index due to {\it Crystals} ($-50$\%), {\it Close Components} ($-60$\%) and {\it Cryostat} ($-60$\%). We also clarify that {\it Close Components} contribution is mainly due to {\it Holder} contaminations, being {\it Reflectors} contaminations at least one order of magnitude lower than those of copper (see Tab.~III~\cite{Supplemental}). In \phtwo this background is higher because of the {\it 10mK} shielding.}

{\it \se-\vv decay half-life.} We fit the \vv spectrum in the background model constraining its activity to be the same in \phone and \phtwo. The reconstructed \se-\vv activity from the background model considering \SeMass of isotope mass is \vvactivitySSD.
Differently from the previous analysis carried out on \phone data~\cite{Azzolini:2019yib}, we include the efficiencies as fit parameters with a Gaussian prior set at their measured value. The resulting marginalized posterior includes the efficiencies and \se isotopic abundance uncertainties as nuisance parameters. Thus, these uncertainties are included in the statistical one ($\pm 0.6 \%$).
The results of the systematic studies are summarized in Fig.~\ref{fig:2nu}, where we report the percentage difference for the \se-\vv half-life between each test and the \textit{Reference} fit. 
For the threshold and the source location, we quote the maximum difference among those of the different tests.
We assume each fit systematics to be single-sided and estimate the uncertainty as 68\% of the difference between the test fit result and the {\it Reference} one. Thus, the overall systematic uncertainty of the \se-\vv half-life is~($^{+1.0\%}_{-0.7\%}$), evaluated as the sum in quadrature of the single uncertainties. In summary, the ultimate value of the \se-\vv half-life value measured by CUPID-0 is
\begin{equation*}
\HalfLifenewred, 
\end{equation*}
fully compatible with the previous measurement reported by CUPID-0, $\HalfLifeold$~\cite{Azzolini:2019yib}.The improved precision can be attributed mainly to the integration of Phase I data with Phase II data. Additionally, the analysis threshold has been lowered from 700 keV to 300 keV, while the uncertainty on the isotopic abundance has been reduced by a factor of three
~\cite{Balabanov:2023tnj}. 
Finally, we can convert the half-life in terms of nuclear matrix element (NME) using the Phase Space Factor $G^{2\nu} = (1.996 \pm 0.028)\times10^{-18}$ y$^{-1}$ obtained under the SSD model~\cite{Kotila:2012zza,JKprivate}. Without making hypotheses on the axial coupling constant $g_A$, we can provide a new experimental value of the effective NME, $\mathcal{M}^{eff}_{2\nu}=\mathcal{M}_{2\nu}g_A^2$~=~$\mvveffred$, to be compared with theoretical calculations. This measurement represents the new benchmark for nuclear models, which currently obtain significantly higher values \cite{Nomura:2022nwv,Barea:2015kwa,Popara:2021lst,Coraggio:2018tuo,Ejiri:2019ezh} and resort to a phenomenological quenching of $g_A$ to explain the discrepancy.

In conclusion, by combining the experimental data collected by CUPID-0 in \phone and \phtwo, we get a better comprehension of the background affecting the region of interest of \se-\onu. Thanks to the excellent data reconstruction, we measure the \se-\vv half-life with an accuracy of $\mathcal{O}(1\%)$, obtaining the most precise measurement of \vv among all isotopes to date~\cite{KamLAND-Zen:2019imh,CUORE:2020bok,CUPID-Mo:2023lru,GERDA:2023wbr}.

\begin{acknowledgments}
This work was partially supported by the European Research Council (FP7/2007-2013), Contract No. 247115. The work of J.~Kotila was supported by the Academy of Finland (Grant Nos. 314733 and 345869).
L. P. research activities are supported by European Union’s Horizon 2020306
research and innovation program under the Marie Skłodowska-Curie grant agreement N. 10102968. We are particularly grateful to S.~Grigioni and M.~Veronelli for their help in the design and construction of the sensor-to-absorber gluing system, M.~Iannone for the help in all the stages of the detector construction, A.~Pelosi for the construction of the assembly line, M.~Guetti for the assistance in the cryogenic operations, R.~Gaigher for the calibration system mechanics, M.~Lindozzi for the development of cryostat monitoring system, M.~Perego for his invaluable help, the mechanical workshop of LNGS (E.~Tatananni, A.~Rotilio, A.~Corsi, and B.~Romualdi) for the continuous help in the overall setup design. We acknowledge the Dark Side Collaboration for the use of the low-radon clean room. This work makes use of the DIANA data analysis and APOLLO data acquisition software which has been developed by the CUORICINO, CUORE, LUCIFER, and CUPID-0 Collaborations. This work makes use of the Arby software for Geant4-based Monte Carlo simulations, which has been developed in the framework of the Milano - Bicocca R\&D activities and that is maintained by O.~Cremonesi and S.~Pozzi.
\end{acknowledgments}
\bibliography{main}

\begin{thebibliography}{49}%
\makeatletter
\providecommand \@ifxundefined [1]{%
 \@ifx{#1\undefined}
}%
\providecommand \@ifnum [1]{%
 \ifnum #1\expandafter \@firstoftwo
 \else \expandafter \@secondoftwo
 \fi
}%
\providecommand \@ifx [1]{%
 \ifx #1\expandafter \@firstoftwo
 \else \expandafter \@secondoftwo
 \fi
}%
\providecommand \natexlab [1]{#1}%
\providecommand \enquote  [1]{``#1''}%
\providecommand \bibnamefont  [1]{#1}%
\providecommand \bibfnamefont [1]{#1}%
\providecommand \citenamefont [1]{#1}%
\providecommand \href@noop [0]{\@secondoftwo}%
\providecommand \href [0]{\begingroup \@sanitize@url \@href}%
\providecommand \@href[1]{\@@startlink{#1}\@@href}%
\providecommand \@@href[1]{\endgroup#1\@@endlink}%
\providecommand \@sanitize@url [0]{\catcode `\\12\catcode `\$12\catcode
  `\&12\catcode `\#12\catcode `\^12\catcode `\_12\catcode `\%12\relax}%
\providecommand \@@startlink[1]{}%
\providecommand \@@endlink[0]{}%
\providecommand \url  [0]{\begingroup\@sanitize@url \@url }%
\providecommand \@url [1]{\endgroup\@href {#1}{\urlprefix }}%
\providecommand \urlprefix  [0]{URL }%
\providecommand \Eprint [0]{\href }%
\providecommand \doibase [0]{http://dx.doi.org/}%
\providecommand \selectlanguage [0]{\@gobble}%
\providecommand \bibinfo  [0]{\@secondoftwo}%
\providecommand \bibfield  [0]{\@secondoftwo}%
\providecommand \translation [1]{[#1]}%
\providecommand \BibitemOpen [0]{}%
\providecommand \bibitemStop [0]{}%
\providecommand \bibitemNoStop [0]{.\EOS\space}%
\providecommand \EOS [0]{\spacefactor3000\relax}%
\providecommand \BibitemShut  [1]{\csname bibitem#1\endcsname}%
\let\auto@bib@innerbib\@empty
%</preamble>
\bibitem [{\citenamefont {Furry}(1939)}]{Furry:1939qr}%
  \BibitemOpen
  \bibfield  {author} {\bibinfo {author} {\bibfnamefont {W.~H.}\ \bibnamefont
  {Furry}},\ }\href {\doibase 10.1103/PhysRev.56.1184} {\bibfield  {journal}
  {\bibinfo  {journal} {Phys. Rev.}\ }\textbf {\bibinfo {volume} {56}},\
  \bibinfo {pages} {1184} (\bibinfo {year} {1939})}\BibitemShut {NoStop}%
%%CITATION = PHRVA,56,1184;%%
\bibitem [{\citenamefont {Dell'Oro}\ \emph {et~al.}(2016)\citenamefont
  {Dell'Oro}, \citenamefont {Marcocci}, \citenamefont {Viel},\ and\
  \citenamefont {Vissani}}]{DellOro:2016tmg}%
  \BibitemOpen
  \bibfield  {author} {\bibinfo {author} {\bibfnamefont {S.}~\bibnamefont
  {Dell'Oro}}, \bibinfo {author} {\bibfnamefont {S.}~\bibnamefont {Marcocci}},
  \bibinfo {author} {\bibfnamefont {M.}~\bibnamefont {Viel}}, \ and\ \bibinfo
  {author} {\bibfnamefont {F.}~\bibnamefont {Vissani}},\ }\href {\doibase
  10.1155/2016/2162659} {\bibfield  {journal} {\bibinfo  {journal} {Adv. High
  Energy Phys.}\ }\textbf {\bibinfo {volume} {2016}},\ \bibinfo {pages}
  {2162659} (\bibinfo {year} {2016})}\BibitemShut {NoStop}%
\bibitem [{\citenamefont {Goeppert-Mayer}(1935)}]{GoeppertMayer}%
  \BibitemOpen
  \bibfield  {author} {\bibinfo {author} {\bibfnamefont {M.}~\bibnamefont
  {Goeppert-Mayer}},\ }\href {\doibase 10.1103/PhysRev.48.512} {\bibfield
  {journal} {\bibinfo  {journal} {Phys. Rev.}\ }\textbf {\bibinfo {volume}
  {48}},\ \bibinfo {pages} {512} (\bibinfo {year} {1935})}\BibitemShut
  {NoStop}%
%%CITATION = PHRVA,48,512;%%
\bibitem [{\citenamefont {Agostini}\ \emph {et~al.}(2020)\citenamefont
  {Agostini} \emph {et~al.}}]{GERDA:2020xhi}%
  \BibitemOpen
  \bibfield  {author} {\bibinfo {author} {\bibfnamefont {M.}~\bibnamefont
  {Agostini}} \emph {et~al.} (\bibinfo {collaboration} {GERDA}),\ }\href
  {\doibase 10.1103/PhysRevLett.125.252502} {\bibfield  {journal} {\bibinfo
  {journal} {Phys. Rev. Lett.}\ }\textbf {\bibinfo {volume} {125}},\ \bibinfo
  {pages} {252502} (\bibinfo {year} {2020})},\ \Eprint
  {http://arxiv.org/abs/2009.06079} {arXiv:2009.06079 [nucl-ex]} \BibitemShut
  {NoStop}%
\bibitem [{\citenamefont {Abe}\ \emph {et~al.}(2023)\citenamefont {Abe} \emph
  {et~al.}}]{KamLAND-Zen:2022tow}%
  \BibitemOpen
  \bibfield  {author} {\bibinfo {author} {\bibfnamefont {S.}~\bibnamefont
  {Abe}} \emph {et~al.} (\bibinfo {collaboration} {KamLAND-Zen}),\ }\href
  {\doibase 10.1103/PhysRevLett.130.051801} {\bibfield  {journal} {\bibinfo
  {journal} {Phys. Rev. Lett.}\ }\textbf {\bibinfo {volume} {130}},\ \bibinfo
  {pages} {051801} (\bibinfo {year} {2023})},\ \Eprint
  {http://arxiv.org/abs/2203.02139} {arXiv:2203.02139 [hep-ex]} \BibitemShut
  {NoStop}%
\bibitem [{\citenamefont {Adams}\ \emph {et~al.}(2022)\citenamefont {Adams}
  \emph {et~al.}}]{CUORE:2021mvw}%
  \BibitemOpen
  \bibfield  {author} {\bibinfo {author} {\bibfnamefont {D.~Q.}\ \bibnamefont
  {Adams}} \emph {et~al.} (\bibinfo {collaboration} {CUORE}),\ }\href {\doibase
  10.1038/s41586-022-04497-4} {\bibfield  {journal} {\bibinfo  {journal}
  {Nature}\ }\textbf {\bibinfo {volume} {604}},\ \bibinfo {pages} {53}
  (\bibinfo {year} {2022})},\ \Eprint {http://arxiv.org/abs/2104.06906}
  {arXiv:2104.06906 [nucl-ex]} \BibitemShut {NoStop}%
\bibitem [{\citenamefont {Saakyan}(2013)}]{Saakyan:2013yna}%
  \BibitemOpen
  \bibfield  {author} {\bibinfo {author} {\bibfnamefont {R.}~\bibnamefont
  {Saakyan}},\ }\href {\doibase 10.1146/annurev-nucl-102711-094904} {\bibfield
  {journal} {\bibinfo  {journal} {Ann. Rev. Nucl. Part. Sci.}\ }\textbf
  {\bibinfo {volume} {63}},\ \bibinfo {pages} {503} (\bibinfo {year}
  {2013})}\BibitemShut {NoStop}%
%%CITATION = ARNUA,63,503;%%
\bibitem [{\citenamefont {Barabash}(2015)}]{Barabash:2015eza}%
  \BibitemOpen
  \bibfield  {author} {\bibinfo {author} {\bibfnamefont {A.~S.}\ \bibnamefont
  {Barabash}},\ }\href {\doibase 10.1016/j.nuclphysa.2015.01.001} {\bibfield
  {journal} {\bibinfo  {journal} {Nucl. Phys.}\ }\textbf {\bibinfo {volume}
  {A935}},\ \bibinfo {pages} {52} (\bibinfo {year} {2015})},\ \Eprint
  {http://arxiv.org/abs/1501.05133} {arXiv:1501.05133 [nucl-ex]} \BibitemShut
  {NoStop}%
%%CITATION = ARXIV:1501.05133;%%
\bibitem [{\citenamefont {Novella}\ \emph {et~al.}(2022)\citenamefont {Novella}
  \emph {et~al.}}]{NEXT:2021dqj}%
  \BibitemOpen
  \bibfield  {author} {\bibinfo {author} {\bibfnamefont {P.}~\bibnamefont
  {Novella}} \emph {et~al.} (\bibinfo {collaboration} {NEXT}),\ }\href
  {\doibase 10.1103/PhysRevC.105.055501} {\bibfield  {journal} {\bibinfo
  {journal} {Phys. Rev. C}\ }\textbf {\bibinfo {volume} {105}},\ \bibinfo
  {pages} {055501} (\bibinfo {year} {2022})},\ \Eprint
  {http://arxiv.org/abs/2111.11091} {arXiv:2111.11091 [nucl-ex]} \BibitemShut
  {NoStop}%
\bibitem [{\citenamefont {Agostini}\ \emph {et~al.}(2023)\citenamefont
  {Agostini} \emph {et~al.}}]{GERDA:2023wbr}%
  \BibitemOpen
  \bibfield  {author} {\bibinfo {author} {\bibfnamefont {M.}~\bibnamefont
  {Agostini}} \emph {et~al.} (\bibinfo {collaboration} {GERDA}),\ }\href
  {\doibase 10.1103/PhysRevLett.131.142501} {\bibfield  {journal} {\bibinfo
  {journal} {Phys. Rev. Lett.}\ }\textbf {\bibinfo {volume} {131}},\ \bibinfo
  {pages} {142501} (\bibinfo {year} {2023})},\ \Eprint
  {http://arxiv.org/abs/2308.09795} {arXiv:2308.09795 [nucl-ex]} \BibitemShut
  {NoStop}%
\bibitem [{\citenamefont {Gando}\ \emph {et~al.}(2019)\citenamefont {Gando}
  \emph {et~al.}}]{KamLAND-Zen:2019imh}%
  \BibitemOpen
  \bibfield  {author} {\bibinfo {author} {\bibfnamefont {A.}~\bibnamefont
  {Gando}} \emph {et~al.} (\bibinfo {collaboration} {KamLAND-Zen}),\ }\href
  {\doibase 10.1103/PhysRevLett.122.192501} {\bibfield  {journal} {\bibinfo
  {journal} {Phys. Rev. Lett.}\ }\textbf {\bibinfo {volume} {122}},\ \bibinfo
  {pages} {192501} (\bibinfo {year} {2019})},\ \Eprint
  {http://arxiv.org/abs/1901.03871} {arXiv:1901.03871 [hep-ex]} \BibitemShut
  {NoStop}%
\bibitem [{\citenamefont {Adams}\ \emph {et~al.}(2021)\citenamefont {Adams}
  \emph {et~al.}}]{CUORE:2020bok}%
  \BibitemOpen
  \bibfield  {author} {\bibinfo {author} {\bibfnamefont {D.~Q.}\ \bibnamefont
  {Adams}} \emph {et~al.} (\bibinfo {collaboration} {CUORE}),\ }\href {\doibase
  10.1103/PhysRevLett.126.171801} {\bibfield  {journal} {\bibinfo  {journal}
  {Phys. Rev. Lett.}\ }\textbf {\bibinfo {volume} {126}},\ \bibinfo {pages}
  {171801} (\bibinfo {year} {2021})},\ \Eprint
  {http://arxiv.org/abs/2012.11749} {arXiv:2012.11749 [nucl-ex]} \BibitemShut
  {NoStop}%
\bibitem [{\citenamefont {Augier}\ \emph {et~al.}(2023)\citenamefont {Augier}
  \emph {et~al.}}]{CUPID-Mo:2023lru}%
  \BibitemOpen
  \bibfield  {author} {\bibinfo {author} {\bibfnamefont {C.}~\bibnamefont
  {Augier}} \emph {et~al.} (\bibinfo {collaboration} {CUPID-Mo}),\ }\href
  {\doibase 10.1103/PhysRevLett.131.162501} {\bibfield  {journal} {\bibinfo
  {journal} {Phys. Rev. Lett.}\ }\textbf {\bibinfo {volume} {131}},\ \bibinfo
  {pages} {162501} (\bibinfo {year} {2023})},\ \Eprint
  {http://arxiv.org/abs/2307.14086} {arXiv:2307.14086 [nucl-ex]} \BibitemShut
  {NoStop}%
\bibitem [{\citenamefont {Arnold}\ \emph {et~al.}(2019)\citenamefont {Arnold}
  \emph {et~al.}}]{NEMO-3:2019gwo}%
  \BibitemOpen
  \bibfield  {author} {\bibinfo {author} {\bibfnamefont {R.}~\bibnamefont
  {Arnold}} \emph {et~al.} (\bibinfo {collaboration} {NEMO-3}),\ }\href
  {\doibase 10.1140/epjc/s10052-019-6948-4} {\bibfield  {journal} {\bibinfo
  {journal} {Eur. Phys. J.}\ }\textbf {\bibinfo {volume} {C79}},\ \bibinfo
  {pages} {440} (\bibinfo {year} {2019})},\ \Eprint
  {http://arxiv.org/abs/1903.08084} {arXiv:1903.08084 [nucl-ex]} \BibitemShut
  {NoStop}%
%%CITATION = ARXIV:1903.08084;%%
\bibitem [{\citenamefont {Pirro}\ \emph {et~al.}(2006)\citenamefont {Pirro}
  \emph {et~al.}}]{Pirro:2005ar}%
  \BibitemOpen
  \bibfield  {author} {\bibinfo {author} {\bibfnamefont {S.}~\bibnamefont
  {Pirro}} \emph {et~al.},\ }\href {\doibase 10.1134/S1063778806120155}
  {\bibfield  {journal} {\bibinfo  {journal} {Phys. Atom. Nucl.}\ }\textbf
  {\bibinfo {volume} {69}},\ \bibinfo {pages} {2109} (\bibinfo {year}
  {2006})}\BibitemShut {NoStop}%
\bibitem [{\citenamefont {Arnaboldi}\ \emph {et~al.}(2011)\citenamefont
  {Arnaboldi}, \citenamefont {Capelli}, \citenamefont {Cremonesi},
  \citenamefont {Gironi}, \citenamefont {Pavan}, \citenamefont {Pessina},\ and\
  \citenamefont {Pirro}}]{Arnaboldi:2010jx}%
  \BibitemOpen
  \bibfield  {author} {\bibinfo {author} {\bibfnamefont {C.}~\bibnamefont
  {Arnaboldi}}, \bibinfo {author} {\bibfnamefont {S.}~\bibnamefont {Capelli}},
  \bibinfo {author} {\bibfnamefont {O.}~\bibnamefont {Cremonesi}}, \bibinfo
  {author} {\bibfnamefont {L.}~\bibnamefont {Gironi}}, \bibinfo {author}
  {\bibfnamefont {M.}~\bibnamefont {Pavan}}, \bibinfo {author} {\bibfnamefont
  {G.}~\bibnamefont {Pessina}}, \ and\ \bibinfo {author} {\bibfnamefont
  {S.}~\bibnamefont {Pirro}},\ }\href {\doibase
  10.1016/j.astropartphys.2010.09.004} {\bibfield  {journal} {\bibinfo
  {journal} {Astropart. Phys.}\ }\textbf {\bibinfo {volume} {34}},\ \bibinfo
  {pages} {344} (\bibinfo {year} {2011})},\ \Eprint
  {http://arxiv.org/abs/1006.2721} {arXiv:1006.2721 [nucl-ex]} \BibitemShut
  {NoStop}%
%%CITATION = ARXIV:1006.2721;%%
\bibitem [{\citenamefont {Beeman}\ \emph
  {et~al.}(2013{\natexlab{a}})\citenamefont {Beeman} \emph
  {et~al.}}]{Beeman:2013vda}%
  \BibitemOpen
  \bibfield  {author} {\bibinfo {author} {\bibfnamefont {J.~W.}\ \bibnamefont
  {Beeman}} \emph {et~al.},\ }\href {\doibase 10.1088/1748-0221/8/05/P05021}
  {\bibfield  {journal} {\bibinfo  {journal} {JINST}\ }\textbf {\bibinfo
  {volume} {8}},\ \bibinfo {pages} {P05021} (\bibinfo {year}
  {2013}{\natexlab{a}})}\BibitemShut {NoStop}%
%%CITATION = ARXIV:1303.4080;%%
\bibitem [{\citenamefont {Artusa}\ \emph {et~al.}(2016)\citenamefont {Artusa}
  \emph {et~al.}}]{Artusa:2016maw}%
  \BibitemOpen
  \bibfield  {author} {\bibinfo {author} {\bibfnamefont {D.~R.}\ \bibnamefont
  {Artusa}} \emph {et~al.},\ }\href {\doibase 10.1140/epjc/s10052-016-4223-5}
  {\bibfield  {journal} {\bibinfo  {journal} {Eur. Phys. J. C}\ }\textbf
  {\bibinfo {volume} {76}},\ \bibinfo {pages} {364} (\bibinfo {year}
  {2016})}\BibitemShut {NoStop}%
\bibitem [{\citenamefont {Armengaud}\ \emph {et~al.}(2017)\citenamefont
  {Armengaud} \emph {et~al.}}]{Armengaud:2017hit}%
  \BibitemOpen
  \bibfield  {author} {\bibinfo {author} {\bibfnamefont {E.}~\bibnamefont
  {Armengaud}} \emph {et~al.},\ }\href {\doibase
  10.1140/epjc/s10052-017-5343-2} {\bibfield  {journal} {\bibinfo  {journal}
  {Eur. Phys. J. C}\ }\textbf {\bibinfo {volume} {77}},\ \bibinfo {pages} {785}
  (\bibinfo {year} {2017})}\BibitemShut {NoStop}%
\bibitem [{\citenamefont {Armstrong}\ \emph {et~al.}(2019)\citenamefont
  {Armstrong} \emph {et~al.}}]{CUPIDInterestGroup:2019inu}%
  \BibitemOpen
  \bibfield  {author} {\bibinfo {author} {\bibfnamefont {W.~R.}\ \bibnamefont
  {Armstrong}} \emph {et~al.} (\bibinfo {collaboration} {CUPID}),\ }\href@noop
  {} {\  (\bibinfo {year} {2019})},\ \Eprint {http://arxiv.org/abs/1907.09376}
  {arXiv:1907.09376 [physics.ins-det]} \BibitemShut {NoStop}%
%%CITATION = ARXIV:1907.09376;%%
\bibitem [{\citenamefont {Dafinei}\ \emph {et~al.}(2017)\citenamefont {Dafinei}
  \emph {et~al.}}]{Dafinei:2017xpc}%
  \BibitemOpen
  \bibfield  {author} {\bibinfo {author} {\bibfnamefont {I.}~\bibnamefont
  {Dafinei}} \emph {et~al.},\ }\href {\doibase 10.1016/j.jcrysgro.2017.06.013}
  {\bibfield  {journal} {\bibinfo  {journal} {J. Cryst. Growth}\ }\textbf
  {\bibinfo {volume} {475}},\ \bibinfo {pages} {158} (\bibinfo {year}
  {2017})},\ \Eprint {http://arxiv.org/abs/1702.05877} {arXiv:1702.05877
  [physics.ins-det]} \BibitemShut {NoStop}%
%%CITATION = ARXIV:1702.05877;%%
\bibitem [{\citenamefont {Balabanov}\ \emph {et~al.}(2023)\citenamefont
  {Balabanov} \emph {et~al.}}]{Balabanov:2023tnj}%
  \BibitemOpen
  \bibfield  {author} {\bibinfo {author} {\bibfnamefont {S.~S.}\ \bibnamefont
  {Balabanov}} \emph {et~al.},\ }\href {\doibase
  10.1088/1748-0221/18/04/P04035} {\bibfield  {journal} {\bibinfo  {journal}
  {JINST}\ }\textbf {\bibinfo {volume} {18}},\ \bibinfo {pages} {P04035}
  (\bibinfo {year} {2023})}\BibitemShut {NoStop}%
\bibitem [{\citenamefont {Beeman}\ \emph
  {et~al.}(2013{\natexlab{b}})\citenamefont {Beeman} \emph
  {et~al.}}]{Beeman:2013zva}%
  \BibitemOpen
  \bibfield  {author} {\bibinfo {author} {\bibfnamefont {J.~W.}\ \bibnamefont
  {Beeman}} \emph {et~al.},\ }\href {\doibase 10.1088/1748-0221/8/07/P07021}
  {\bibfield  {journal} {\bibinfo  {journal} {JINST}\ }\textbf {\bibinfo
  {volume} {8}},\ \bibinfo {pages} {P07021} (\bibinfo {year}
  {2013}{\natexlab{b}})},\ \Eprint {http://arxiv.org/abs/1304.6289}
  {arXiv:1304.6289 [physics.ins-det]} \BibitemShut {NoStop}%
%%CITATION = ARXIV:1304.6289;%%
\bibitem [{\citenamefont {Azzolini}\ \emph
  {et~al.}(2018{\natexlab{a}})\citenamefont {Azzolini} \emph
  {et~al.}}]{Azzolini:2018tum}%
  \BibitemOpen
  \bibfield  {author} {\bibinfo {author} {\bibfnamefont {O.}~\bibnamefont
  {Azzolini}} \emph {et~al.} (\bibinfo {collaboration} {CUPID}),\ }\href
  {\doibase 10.1140/epjc/s10052-018-5896-8} {\bibfield  {journal} {\bibinfo
  {journal} {Eur. Phys. J.}\ }\textbf {\bibinfo {volume} {C78}},\ \bibinfo
  {pages} {428} (\bibinfo {year} {2018}{\natexlab{a}})},\ \Eprint
  {http://arxiv.org/abs/1802.06562} {arXiv:1802.06562 [physics.ins-det]}
  \BibitemShut {NoStop}%
%%CITATION = ARXIV:1802.06562;%%
\bibitem [{\citenamefont {Alfonso}\ \emph {et~al.}(2018)\citenamefont
  {Alfonso}, \citenamefont {Cassina}, \citenamefont {Giachero}, \citenamefont
  {Gotti}, \citenamefont {Pessina},\ and\ \citenamefont
  {Carniti}}]{Carniti:2017zkr}%
  \BibitemOpen
  \bibfield  {author} {\bibinfo {author} {\bibfnamefont {K.}~\bibnamefont
  {Alfonso}}, \bibinfo {author} {\bibfnamefont {L.}~\bibnamefont {Cassina}},
  \bibinfo {author} {\bibfnamefont {A.}~\bibnamefont {Giachero}}, \bibinfo
  {author} {\bibfnamefont {C.}~\bibnamefont {Gotti}}, \bibinfo {author}
  {\bibfnamefont {G.}~\bibnamefont {Pessina}}, \ and\ \bibinfo {author}
  {\bibfnamefont {P.}~\bibnamefont {Carniti}},\ }\href {\doibase
  10.1088/1748-0221/13/02/P02029} {\bibfield  {journal} {\bibinfo  {journal}
  {JINST}\ }\textbf {\bibinfo {volume} {13}},\ \bibinfo {pages} {P02029}
  (\bibinfo {year} {2018})},\ \Eprint {http://arxiv.org/abs/1710.05565}
  {arXiv:1710.05565 [physics.ins-det]} \BibitemShut {NoStop}%
\bibitem [{\citenamefont {Arnaboldi}\ \emph {et~al.}(2015)\citenamefont
  {Arnaboldi} \emph {et~al.}}]{Arnaboldi:2015wvc}%
  \BibitemOpen
  \bibfield  {author} {\bibinfo {author} {\bibfnamefont {C.}~\bibnamefont
  {Arnaboldi}} \emph {et~al.},\ }\href {\doibase 10.1063/1.4936269} {\bibfield
  {journal} {\bibinfo  {journal} {Rev. Sci. Instrum.}\ }\textbf {\bibinfo
  {volume} {86}},\ \bibinfo {pages} {124703} (\bibinfo {year}
  {2015})}\BibitemShut {NoStop}%
%%CITATION = RSINA,86,124703;%%
\bibitem [{\citenamefont {Arnaboldi}\ \emph {et~al.}(2018)\citenamefont
  {Arnaboldi} \emph {et~al.}}]{Arnaboldi:2017aek}%
  \BibitemOpen
  \bibfield  {author} {\bibinfo {author} {\bibfnamefont {C.}~\bibnamefont
  {Arnaboldi}} \emph {et~al.},\ }\href {\doibase
  10.1088/1748-0221/13/02/P02026} {\bibfield  {journal} {\bibinfo  {journal}
  {JINST}\ }\textbf {\bibinfo {volume} {13}},\ \bibinfo {pages} {P02026}
  (\bibinfo {year} {2018})},\ \Eprint {http://arxiv.org/abs/1710.06365}
  {arXiv:1710.06365 [physics.ins-det]} \BibitemShut {NoStop}%
%%CITATION = ARXIV:1710.06365;%%
\bibitem [{\citenamefont {Azzolini}\ \emph {et~al.}(2021)\citenamefont
  {Azzolini} \emph {et~al.}}]{Azzolini:2021yft}%
  \BibitemOpen
  \bibfield  {author} {\bibinfo {author} {\bibfnamefont {O.}~\bibnamefont
  {Azzolini}} \emph {et~al.},\ }\href {\doibase
  10.1140/epjc/s10052-021-09476-z} {\bibfield  {journal} {\bibinfo  {journal}
  {Eur. Phys. J. C}\ }\textbf {\bibinfo {volume} {81}},\ \bibinfo {pages} {722}
  (\bibinfo {year} {2021})},\ \Eprint {http://arxiv.org/abs/2105.04409}
  {arXiv:2105.04409 [physics.ins-det]} \BibitemShut {NoStop}%
\bibitem [{\citenamefont {Armatol}\ \emph {et~al.}(2021)\citenamefont {Armatol}
  \emph {et~al.}}]{CUPID:2020itw}%
  \BibitemOpen
  \bibfield  {author} {\bibinfo {author} {\bibfnamefont {A.}~\bibnamefont
  {Armatol}} \emph {et~al.} (\bibinfo {collaboration} {CUPID}),\ }\href
  {\doibase 10.1140/epjc/s10052-020-08809-8} {\bibfield  {journal} {\bibinfo
  {journal} {Eur. Phys. J. C}\ }\textbf {\bibinfo {volume} {81}},\ \bibinfo
  {pages} {104} (\bibinfo {year} {2021})},\ \Eprint
  {http://arxiv.org/abs/2011.13656} {arXiv:2011.13656 [physics.ins-det]}
  \BibitemShut {NoStop}%
\bibitem [{\citenamefont {Azzolini}\ \emph {et~al.}(2022)\citenamefont
  {Azzolini} \emph {et~al.}}]{CUPID:2022puj}%
  \BibitemOpen
  \bibfield  {author} {\bibinfo {author} {\bibfnamefont {O.}~\bibnamefont
  {Azzolini}} \emph {et~al.} (\bibinfo {collaboration} {CUPID}),\ }\href
  {\doibase 10.1103/PhysRevLett.129.111801} {\bibfield  {journal} {\bibinfo
  {journal} {Phys. Rev. Lett.}\ }\textbf {\bibinfo {volume} {129}},\ \bibinfo
  {pages} {111801} (\bibinfo {year} {2022})},\ \Eprint
  {http://arxiv.org/abs/2206.05130} {arXiv:2206.05130 [nucl-ex]} \BibitemShut
  {NoStop}%
\bibitem [{\citenamefont {Di~Domizio}\ \emph {et~al.}(2018)\citenamefont
  {Di~Domizio} \emph {et~al.}}]{DiDomizio:2018ldc}%
  \BibitemOpen
  \bibfield  {author} {\bibinfo {author} {\bibfnamefont {S.}~\bibnamefont
  {Di~Domizio}} \emph {et~al.},\ }\href {\doibase
  10.1088/1748-0221/13/12/P12003} {\bibfield  {journal} {\bibinfo  {journal}
  {JINST}\ }\textbf {\bibinfo {volume} {13}},\ \bibinfo {pages} {P12003}
  (\bibinfo {year} {2018})},\ \Eprint {http://arxiv.org/abs/1807.11446}
  {arXiv:1807.11446 [physics.ins-det]} \BibitemShut {NoStop}%
%%CITATION = ARXIV:1807.11446;%%
\bibitem [{\citenamefont {Gatti}\ and\ \citenamefont
  {Manfredi}(1986)}]{Gatti:1986cw}%
  \BibitemOpen
  \bibfield  {author} {\bibinfo {author} {\bibfnamefont {E.}~\bibnamefont
  {Gatti}}\ and\ \bibinfo {author} {\bibfnamefont {P.~F.}\ \bibnamefont
  {Manfredi}},\ }\href@noop {} {\bibfield  {journal} {\bibinfo  {journal} {Riv.
  Nuovo Cimento}\ }\textbf {\bibinfo {volume} {9}},\ \bibinfo {pages} {1}
  (\bibinfo {year} {1986})}\BibitemShut {NoStop}%
%%CITATION = RNCIB,9N1,1;%%
\bibitem [{\citenamefont {Azzolini}\ \emph
  {et~al.}(2019{\natexlab{a}})\citenamefont {Azzolini} \emph
  {et~al.}}]{Azzolini:2019tta}%
  \BibitemOpen
  \bibfield  {author} {\bibinfo {author} {\bibfnamefont {O.}~\bibnamefont
  {Azzolini}} \emph {et~al.} (\bibinfo {collaboration} {CUPID}),\ }\href
  {\doibase 10.1103/PhysRevLett.123.032501} {\bibfield  {journal} {\bibinfo
  {journal} {Phys. Rev. Lett.}\ }\textbf {\bibinfo {volume} {123}},\ \bibinfo
  {pages} {032501} (\bibinfo {year} {2019}{\natexlab{a}})},\ \Eprint
  {http://arxiv.org/abs/1906.05001} {arXiv:1906.05001 [nucl-ex]} \BibitemShut
  {NoStop}%
%%CITATION = ARXIV:1906.05001;%%
\bibitem [{\citenamefont {Azzolini}\ \emph
  {et~al.}(2018{\natexlab{b}})\citenamefont {Azzolini} \emph
  {et~al.}}]{Azzolini:2018yye}%
  \BibitemOpen
  \bibfield  {author} {\bibinfo {author} {\bibfnamefont {O.}~\bibnamefont
  {Azzolini}} \emph {et~al.},\ }\href {\doibase 10.1140/epjc/s10052-018-6202-5}
  {\bibfield  {journal} {\bibinfo  {journal} {Eur. Phys. J.}\ }\textbf
  {\bibinfo {volume} {C78}},\ \bibinfo {pages} {734} (\bibinfo {year}
  {2018}{\natexlab{b}})},\ \Eprint {http://arxiv.org/abs/1806.02826}
  {arXiv:1806.02826 [physics.ins-det]} \BibitemShut {NoStop}%
%%CITATION = ARXIV:1806.02826;%%
\bibitem [{\citenamefont {Azzolini}\ \emph
  {et~al.}(2019{\natexlab{b}})\citenamefont {Azzolini} \emph
  {et~al.}}]{Azzolini:2019nmi}%
  \BibitemOpen
  \bibfield  {author} {\bibinfo {author} {\bibfnamefont {O.}~\bibnamefont
  {Azzolini}} \emph {et~al.} (\bibinfo {collaboration} {CUPID}),\ }\href
  {\doibase 10.1140/epjc/s10052-019-7078-8} {\bibfield  {journal} {\bibinfo
  {journal} {Eur. Phys. J.}\ }\textbf {\bibinfo {volume} {C79}},\ \bibinfo
  {pages} {583} (\bibinfo {year} {2019}{\natexlab{b}})},\ \Eprint
  {http://arxiv.org/abs/1904.10397} {arXiv:1904.10397 [nucl-ex]} \BibitemShut
  {NoStop}%
%%CITATION = ARXIV:1904.10397;%%
\bibitem [{\citenamefont {Agostinelli}\ \emph {et~al.}(2003)\citenamefont
  {Agostinelli} \emph {et~al.}}]{Geant4}%
  \BibitemOpen
  \bibfield  {author} {\bibinfo {author} {\bibfnamefont {S.}~\bibnamefont
  {Agostinelli}} \emph {et~al.},\ }\href {\doibase
  10.1016/S0168-9002(03)01368-8} {\bibfield  {journal} {\bibinfo  {journal}
  {Nucl. Instr. Meth. A}\ }\textbf {\bibinfo {volume} {506}},\ \bibinfo {pages}
  {250} (\bibinfo {year} {2003})}\BibitemShut {NoStop}%
\bibitem [{\citenamefont {Kotila}\ and\ \citenamefont
  {Iachello}(2012)}]{Kotila:2012zza}%
  \BibitemOpen
  \bibfield  {author} {\bibinfo {author} {\bibfnamefont {J.}~\bibnamefont
  {Kotila}}\ and\ \bibinfo {author} {\bibfnamefont {F.}~\bibnamefont
  {Iachello}},\ }\href {\doibase 10.1103/PhysRevC.85.034316} {\bibfield
  {journal} {\bibinfo  {journal} {Phys. Rev.}\ }\textbf {\bibinfo {volume}
  {C85}},\ \bibinfo {pages} {034316} (\bibinfo {year} {2012})},\ \Eprint
  {http://arxiv.org/abs/1209.5722} {arXiv:1209.5722 [nucl-th]} \BibitemShut
  {NoStop}%
%%CITATION = ARXIV:1209.5722;%%
\bibitem [{\citenamefont {Andreotti}\ \emph {et~al.}(2010)\citenamefont
  {Andreotti} \emph {et~al.}}]{CUORICINO:2009zeh}%
  \BibitemOpen
  \bibfield  {author} {\bibinfo {author} {\bibfnamefont {E.}~\bibnamefont
  {Andreotti}} \emph {et~al.} (\bibinfo {collaboration} {CUORICINO}),\ }\href
  {\doibase 10.1016/j.astropartphys.2010.04.004} {\bibfield  {journal}
  {\bibinfo  {journal} {Astropart. Phys.}\ }\textbf {\bibinfo {volume} {34}},\
  \bibinfo {pages} {18} (\bibinfo {year} {2010})},\ \Eprint
  {http://arxiv.org/abs/0912.3779} {arXiv:0912.3779 [nucl-ex]} \BibitemShut
  {NoStop}%
\bibitem [{\citenamefont {Alduino}\ \emph {et~al.}(2017)\citenamefont {Alduino}
  \emph {et~al.}}]{Alduino:2016vtd}%
  \BibitemOpen
  \bibfield  {author} {\bibinfo {author} {\bibfnamefont {C.}~\bibnamefont
  {Alduino}} \emph {et~al.} (\bibinfo {collaboration} {CUORE}),\ }\href
  {\doibase 10.1140/epjc/s10052-016-4498-6} {\bibfield  {journal} {\bibinfo
  {journal} {Eur. Phys. J.}\ }\textbf {\bibinfo {volume} {C77}},\ \bibinfo
  {pages} {13} (\bibinfo {year} {2017})},\ \Eprint
  {http://arxiv.org/abs/1609.01666} {arXiv:1609.01666 [nucl-ex]} \BibitemShut
  {NoStop}%
%%CITATION = ARXIV:1609.01666;%%
\bibitem [{\citenamefont {Ambrosio}\ \emph {et~al.}(1995)\citenamefont
  {Ambrosio} \emph {et~al.}}]{Ambrosio:1995cx}%
  \BibitemOpen
  \bibfield  {author} {\bibinfo {author} {\bibfnamefont {M.}~\bibnamefont
  {Ambrosio}} \emph {et~al.} (\bibinfo {collaboration} {MACRO}),\ }\href
  {\doibase 10.1103/PhysRevD.52.3793} {\bibfield  {journal} {\bibinfo
  {journal} {Phys. Rev.}\ }\textbf {\bibinfo {volume} {D52}},\ \bibinfo {pages}
  {3793} (\bibinfo {year} {1995})}\BibitemShut {NoStop}%
%%CITATION = PHRVA,D52,3793;%%
\bibitem [{\citenamefont {Plummer}(2003)}]{JAGS}%
  \BibitemOpen
  \bibfield  {author} {\bibinfo {author} {\bibfnamefont {M.}~\bibnamefont
  {Plummer}},\ }\href@noop {} {\bibfield  {journal} {\bibinfo  {journal} {JAGS:
  A Program for Analysis of Bayesian Graphical Models using Gibbs Sampling. 3rd
  International Workshop on Distributed Statistical Computing (DSC 2003);
  Vienna, Austria.}\ }\textbf {\bibinfo {volume} {124}} (\bibinfo {year}
  {2003})}\BibitemShut {NoStop}%
\bibitem [{Sup()}]{Supplemental}%
  \BibitemOpen
  \href@noop {} {}\bibinfo {howpublished} {See Supplemental Material [url] for
  further details.}\BibitemShut {Stop}%
\bibitem [{\citenamefont {Azzolini}\ \emph
  {et~al.}(2019{\natexlab{c}})\citenamefont {Azzolini} \emph
  {et~al.}}]{Azzolini:2019yib}%
  \BibitemOpen
  \bibfield  {author} {\bibinfo {author} {\bibfnamefont {O.}~\bibnamefont
  {Azzolini}} \emph {et~al.},\ }\href {\doibase 10.1103/PhysRevLett.123.262501}
  {\bibfield  {journal} {\bibinfo  {journal} {Phys. Rev. Lett.}\ }\textbf
  {\bibinfo {volume} {123}},\ \bibinfo {pages} {262501} (\bibinfo {year}
  {2019}{\natexlab{c}})},\ \Eprint {http://arxiv.org/abs/1909.03397}
  {arXiv:1909.03397 [nucl-ex]} \BibitemShut {NoStop}%
\bibitem [{\citenamefont {Kotila}(2023)}]{JKprivate}%
  \BibitemOpen
  \bibfield  {author} {\bibinfo {author} {\bibfnamefont {J.}~\bibnamefont
  {Kotila}},\ }\href@noop {} {}\bibinfo {howpublished} {Private communication}
  (\bibinfo {year} {2023})\BibitemShut {NoStop}%
\bibitem [{\citenamefont {Nomura}(2022)}]{Nomura:2022nwv}%
  \BibitemOpen
  \bibfield  {author} {\bibinfo {author} {\bibfnamefont {K.}~\bibnamefont
  {Nomura}},\ }\href {\doibase 10.1103/PhysRevC.105.044301} {\bibfield
  {journal} {\bibinfo  {journal} {Phys. Rev. C}\ }\textbf {\bibinfo {volume}
  {105}},\ \bibinfo {pages} {044301} (\bibinfo {year} {2022})},\ \Eprint
  {http://arxiv.org/abs/2203.13042} {arXiv:2203.13042 [nucl-th]} \BibitemShut
  {NoStop}%
\bibitem [{\citenamefont {Barea}\ \emph {et~al.}(2015)\citenamefont {Barea},
  \citenamefont {Kotila},\ and\ \citenamefont {Iachello}}]{Barea:2015kwa}%
  \BibitemOpen
  \bibfield  {author} {\bibinfo {author} {\bibfnamefont {J.}~\bibnamefont
  {Barea}}, \bibinfo {author} {\bibfnamefont {J.}~\bibnamefont {Kotila}}, \
  and\ \bibinfo {author} {\bibfnamefont {F.}~\bibnamefont {Iachello}},\ }\href
  {\doibase 10.1103/PhysRevC.91.034304} {\bibfield  {journal} {\bibinfo
  {journal} {Phys. Rev.}\ }\textbf {\bibinfo {volume} {C91}},\ \bibinfo {pages}
  {034304} (\bibinfo {year} {2015})},\ \Eprint
  {http://arxiv.org/abs/1506.08530} {arXiv:1506.08530 [nucl-th]} \BibitemShut
  {NoStop}%
%%CITATION = ARXIV:1506.08530;%%
\bibitem [{\citenamefont {Popara}\ \emph {et~al.}(2022)\citenamefont {Popara},
  \citenamefont {Ravli\'c},\ and\ \citenamefont {Paar}}]{Popara:2021lst}%
  \BibitemOpen
  \bibfield  {author} {\bibinfo {author} {\bibfnamefont {N.}~\bibnamefont
  {Popara}}, \bibinfo {author} {\bibfnamefont {A.}~\bibnamefont {Ravli\'c}}, \
  and\ \bibinfo {author} {\bibfnamefont {N.}~\bibnamefont {Paar}},\ }\href
  {\doibase 10.1103/PhysRevC.105.064315} {\bibfield  {journal} {\bibinfo
  {journal} {Phys. Rev. C}\ }\textbf {\bibinfo {volume} {105}},\ \bibinfo
  {pages} {064315} (\bibinfo {year} {2022})},\ \Eprint
  {http://arxiv.org/abs/2107.08747} {arXiv:2107.08747 [nucl-th]} \BibitemShut
  {NoStop}%
\bibitem [{\citenamefont {Coraggio}\ \emph {et~al.}(2019)\citenamefont
  {Coraggio}, \citenamefont {De~Angelis}, \citenamefont {Fukui}, \citenamefont
  {Gargano}, \citenamefont {Itaco},\ and\ \citenamefont
  {Nowacki}}]{Coraggio:2018tuo}%
  \BibitemOpen
  \bibfield  {author} {\bibinfo {author} {\bibfnamefont {L.}~\bibnamefont
  {Coraggio}}, \bibinfo {author} {\bibfnamefont {L.}~\bibnamefont
  {De~Angelis}}, \bibinfo {author} {\bibfnamefont {T.}~\bibnamefont {Fukui}},
  \bibinfo {author} {\bibfnamefont {A.}~\bibnamefont {Gargano}}, \bibinfo
  {author} {\bibfnamefont {N.}~\bibnamefont {Itaco}}, \ and\ \bibinfo {author}
  {\bibfnamefont {F.}~\bibnamefont {Nowacki}},\ }\href {\doibase
  10.1103/PhysRevC.100.014316} {\bibfield  {journal} {\bibinfo  {journal}
  {Phys. Rev. C}\ }\textbf {\bibinfo {volume} {100}},\ \bibinfo {pages}
  {014316} (\bibinfo {year} {2019})},\ \Eprint
  {http://arxiv.org/abs/1812.04292} {arXiv:1812.04292 [nucl-th]} \BibitemShut
  {NoStop}%
\bibitem [{\citenamefont {Ejiri}\ \emph {et~al.}(2019)\citenamefont {Ejiri},
  \citenamefont {Suhonen},\ and\ \citenamefont {Zuber}}]{Ejiri:2019ezh}%
  \BibitemOpen
  \bibfield  {author} {\bibinfo {author} {\bibfnamefont {H.}~\bibnamefont
  {Ejiri}}, \bibinfo {author} {\bibfnamefont {J.}~\bibnamefont {Suhonen}}, \
  and\ \bibinfo {author} {\bibfnamefont {K.}~\bibnamefont {Zuber}},\ }\href
  {\doibase 10.1016/j.physrep.2018.12.001} {\bibfield  {journal} {\bibinfo
  {journal} {Phys. Rept.}\ }\textbf {\bibinfo {volume} {797}},\ \bibinfo
  {pages} {1} (\bibinfo {year} {2019})}\BibitemShut {NoStop}%
\end{thebibliography}%
\end{document}